\documentclass[12pt]{article}
\usepackage{latexsym}
\usepackage{amssymb}
\usepackage{amsmath}
\usepackage{graphicx}
\newcommand{\pme}{^{\prime}}
\newcommand{\gam}{\gamma(x,t)}
\newcommand{\lam}{\lambda(x,t)}

\newcommand{\C}{8\pi}

\newcommand{\at}{a(t)}
\newcommand{\p}{P(x)}
\newcommand{\ftr}{\left[1+P(x)_{,x}^{2}\right]}

\newcommand{\taux}{8\pi T^{x}_{\; x}-\frac{2}{D-2}\Lambda}
\newcommand{\tauxp}{8\pi T^{x}_{\; x}(x^{\prime})-\frac{2}{D-2}\Lambda}

\begin{document}

\title{Higher Dimensional Wormhole Geometries with Compact Dimensions}
\author {{\small A. DeBenedictis \footnote{e-mail: adebened@langara.bc.ca}} \\
\it{\small Department of Physics} \\
\it{\small Langara College, 100 W. 49$^{\mbox{th}}$ ave.,
Vancouver, British Columbia, Canada V5Y 2Z6} \and {\small A. Das \footnote{e-mail: das@sfu.ca}} \\
\it{\small Department of Mathematics} \\ \it{\small Simon Fraser
University, Burnaby, British Columbia, Canada V5A 1S6}}
\date{{\small December 5, 2002}}
\maketitle

\begin{abstract}
This paper studies wormhole solutions to Einstein gravity with an
arbitrary number of time dependent compact dimensions and a
matter-vacuum boundary. A new gauge is utilized which is
particularly suited for studies of the wormhole throat. The
solutions possess arbitrary functions which allow for the
description of infinitely many wormhole systems of this type and,
at the stellar boundary, the matter field is smoothly joined to
vacuum. It turns out that the classical vacuum structure differs
considerably from the four dimensional theory and is therefore
studied in detail. The presence of the vacuum-matter boundary and
extra dimensions places interesting restrictions on the wormhole.
For example, in the static case, the radial size of a weak energy
condition (WEC) respecting throat is restricted by the extra
dimensions. There is a critical dimension, $D=5$, where this
restriction is eliminated. In the time dependent case, one
\emph{cannot} respect the WEC at the throat as the time
dependence actually tends the solution towards WEC violation.
This differs considerably from the static case and the four
dimensional case.
\end{abstract}

\vspace{3mm}

\section{Introduction}
Spacetimes with non-trivial topology have been of interest at
least since the classic works of Flamm \cite{ref:flamm},
 Einstein and Rosen \cite{ref:einstros} and Weyl \cite{ref:weyl1}.
The Einstein-Rosen bridge, for example, connected two sheets and was to
represent an elementary particle. Later, in studies of quantum
gravity, the wormhole was used as a model for the spacetime ``foam''
\cite{ref:wheelfoam} - \cite{ref:garat} which manifests itself at
energies near the Planck scale.

\qquad The past decade has produced many interesting studies of
Lorentzian as well as Euclidean wormholes (much of this work is
summarized in the book by Visser \cite{ref:visbook} and references
therein). This resurgence was mainly due to the meticulous study by
Morris and Thorne \cite{ref:morthorn} of static, spherically symmetric
wormholes and their traversability properties. Such studies of
non-trivial topologies have widespread applicability in studies such
as chronology protection \cite{ref:hawkchron}, topology change
\cite{ref:topchange}, as well as horizon and singularity structure
\cite{ref:fronov}, \cite{ref:vishoc}. For example, it may be possible
that semiclassical effects in high curvature regions may cause wormhole
formation and, via this mechanism, avoid a singularity. This would
have much relevance to the information loss problem. Wormhole
structures have also been considered in the field of cosmic strings
\cite{ref:cosstring}. Interestingly, the wormhole may not only be
limited to the domain of the theorist. Some very interesting studies
have been performed regarding possible astrophysical signatures of
such objects \cite{ref:tor1} - \cite{ref:tor4}.

\qquad Theories of unification (for example, superstring theories)
tend to require extra spatial dimensions to be consistent. The
gravitational sector of these theories usually reproduce higher
dimensional general relativity at energies below the Planck scale, as
is studied here. Lacking a fully consistent theory of quantum gravity,
it is of interest to ask what the physical requirements of systems are
at the classical level. In the semi-classical regime, our study of the
stress-energy tensor will dictate what properties the expectation
value, $\langle\Psi| T^{\mu}_{\;\nu}|\Psi\rangle$, must possess in
order to support a higher dimensional non-trivial topology.

\qquad Many gravitational studies have been performed taking into
account the possibility of extra dimensions (for example
\cite{ref:highdim1} - \cite{ref:debnath}). Often, however, the
extra dimensions considered are non-compact and all dimensions are
considered equivalent. Here we are concerned with the more
realistic case of Kaluza-Klein type extra dimensions. Theories to
explain high-energy phenomena even allow, or require, the
possibility of large extra spacetime dimensions
\cite{ref:arkani}-\cite{ref:largeextra}. In some of these
theories, the standard model fields may freely propagate in the
higher dimensions when their energies are above a relatively low
(roughly electro-weak) scale \cite{ref:arkani} such as would be
found in the early (yet sub-Planckian) universe or in high-energy
astrophysical systems or colliders. We are therefore not
necessarily interested solely in wormholes traversable by large
objects as is often studied in the literature. Finally, higher
dimensional relativity is often studied to discover if any
interesting features are revealed in dimensions other than four.
This avenue has proved fruitful in the case of lower dimensional
black holes \cite{ref:btz} as well as higher dimensional stars
\cite{ref:pdl}, \cite{ref:dasdeb}. It is therefore of interest to
include the possibility of higher dimensions in studies of
wormholes.

\qquad The above motivations indicate that it is instructive to study
wormhole systems of a general class. We attempt to make the models here
physically reasonable and therefore work with the matter field (the
stress-energy tensor) whenever possible. In the next section we discuss
a gauge (coordinate system) which we briefly introduced in
\cite{ref:debdas}. This gauge offers advantages when studying the
near-throat region over the usual coordinates employed in the study of
wormholes (such as the proper-radius and curvature gauges). In this
gauge, we calculate the {\em general} Einstein tensor, orthonormal
Riemann components and conservation law for time-dependent wormholes
with an arbitrary number of compact dimensions. These equations may
also be utilized for non-wormhole systems.

\qquad It is a well known fact that, with the four dimensional
static wormhole, one must violate the weak energy condition (WEC)
at least in some small region near (but not necessarily at) the
throat \cite{ref:morthornyurt}, \cite{ref:kuhfittig},
\cite{ref:debdas}. It has also been shown that, if one drops the
staticity requirement, the WEC violation may be avoided at least
for certain intervals of time \cite{ref:kar} - \cite{ref:li} for
scale factors without a boundary. The null energy condition (NEC)
has also been discussed \cite{ref:hocvispens},
\cite{ref:hocvisnull}, \cite{ref:hayward}. In section
\ref{subsec:quasistat} we construct a general class of wormhole
geometries which are static in the four non-compact dimensions
but may possess time dependent compact extra dimensions. That is,
we consider higher dimensional versions of the well studied four
dimensional static systems. These are physically relevant if one
wishes to consider wormholes in a universe where the expansion of
the non-compact dimensions is negligible, as in studies of stars.
However, unlike the four dimensional case, it is shown that the
WEC at the throat either places restrictions on the wormhole's
size or cannot be satisfied, depending on the model. This is a
direct consequence of the presence a stellar-vacuum boundary and
the extra dimensions. Energy condition analysis is limited to the
near-throat region since, away from  the throat, it may be
possible to patch the wormhole to energy condition respecting
matter. We examine the energy conditions for the non-diagonal
stress-energy tensor as it turns out that for $D>4$, one may
demand staticity in the non-compact dimensions and still possess
a non-zero radial energy flux. We concentrate on the weak energy
condition since, of the well known energy conditions, it is the
easiest to satisfy and it may be argued that the energy
conditions are too restrictive a criterion in any case
\cite{ref:bracvis}. That is, their violation seems to show up in
acceptable {\em classical} as well as quantum field theories and
therefore do not imply that energy condition violating systems
are necessarily unphysical.

\qquad To make the solutions as physically relevant as possible,
we patch the ``star'' supporting the non-trivial topology to the
vacuum. Doing so requires us to solve the vacuum field equations.
(We define the classical vacuum as one which satisfies
$G^{\mu}_{\;\nu}= \Lambda \,\delta^{\mu}_{\;\nu}$
\footnote{Conventions follow those of \cite{ref:MTW} with
$G_{D}$, the $D$-dimensional Newton's constant, and $c \equiv 1$.
Here, Greek indices take on values $0 \rightarrow D-1$ whereas
lower-case Latin indices take on values $1 \rightarrow D-1$.
Indices referring solely to the compact dimensions will be
denoted by capital Latin letters.}). It turns out that, with
compact extra dimensions, a unique spherically symmetric vacuum
{\em does not exist} and we discuss each allowable vacuum in some
detail. Some vacua are even space {\em and} time dependent and
therefore there is \emph{no} Birkhoff's theorem with compact
dimensions. Admittedly, some vacua may be ruled out by
experiment. These ``bounded'' objects may possess greater
astrophysical relevance than those of infinite extent and shed
light on how the presence of a second boundary affects the
interior structure. We also admit non-zero cosmological constant
as will be required for the vacuum product manifold. This is not
only relevant from a cosmological perspective
\cite{ref:cosconst1}, \cite{ref:cosconst2} but yields a much
richer vacuum structure as will be discussed in section
\ref{subsec:vacstruct} below. We concentrate on making the
solutions event horizon and singularity free as there is much
interest in the literature in the possibility of a wormhole being
traversable. (Even if one is considering high energy wormholes
such as those manifest in the early universe, traversability may
be important for particle transport which may lead to the large
scale homogeneity of the universe we observe today.)

\section{Geometry and topology} \label{sec:getop}
\begin{figure}[ht]
\begin{center}
\includegraphics[bb=45 174 492 489, clip, scale=0.5, keepaspectratio=true]{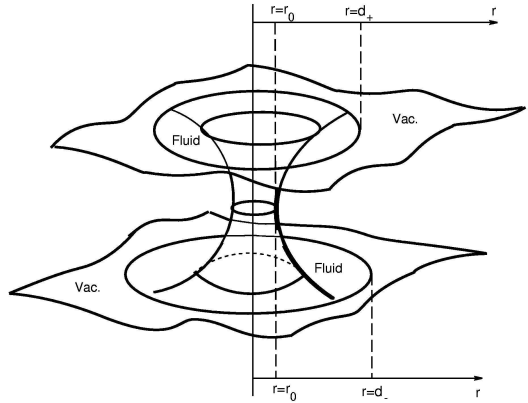}
\caption{{\small A $t=\mbox{constant}$ embedding diagram for the
wormhole geometry. There is a throat at $r=r_{0}$ and stellar
boundaries at $r=d_{\pm}$ where the system must smoothly join the
vacuum.}} \label{fig:1}
\end{center}
\end{figure}

We study here the spherical geometry which is qualitatively
represented by figure \ref{fig:1} which displays a constant time
slice with all but one angle suppressed. The wormhole possesses a
throat of radius $r=r_{0}$ and may connect two regions
representing different universes or else connect two distant
parts of the same universe. The matter boundary is located at
$r=d_{+}$ and $r=d_{-}$  ($+$ and $-$ indicate the upper sheet
and lower sheet respectively) where the interior solution joins
the vacuum. An in-depth analysis of such a ``stellar system''
which is four dimensional and explicitly static may be found in
\cite{ref:debdas}.

\qquad In the curvature coordinates, labeled in figure \ref{fig:1}, one
usually employs the following line element:
\begin{equation}
ds^{2}=-e^{\gamma(r,t)_{\pm}}\,dt^{2}+ e^{\alpha(r,t)_{\pm}}\, dr^{2}
+r^{2}\,d\theta^{2} + r^{2}\sin^{2}(\theta)\,d\phi^{2} +
a(t)\,d\Omega^{2}_{(D-4)}, \label{eq:curvline}
\end{equation}
where we have included the possibility of extra dimensions via
$d\Omega^{2}_{(D-4)}$ thus yielding a warped product manifold of
$M^{4}\times S^{D-4}$. The line element on a unit $D-4$ sphere is:
\begin{equation}
d\Omega^{2}_{(D-4)}=\left[d\psi_{(0)}^{2}+\sum_{n=1}^{D-5}d\psi_{(n)}^{2}
\left(\prod_{m=1}^{n}\sin^{2}\psi_{(m-1)}\right)\right]. \nonumber
\end{equation}

\qquad This chart, though useful, is not ideal for analysis of the
throat. First, there is a coordinate singularity at $r=r_{0}$. Also, an
explicit junction occurs here in these coordinates which must be
addressed via the introduction of an ``upper'' chart and a ``lower''
chart (the $+$ and $-$ indicated above). Therefore, one must carefully
consider junction conditions at the throat to possess an acceptable
patching from the upper to lower chart.

\qquad Alternately, another common chart is the proper radius chart
whose line element is given by:
\begin{equation}
ds^{2}=-e^{\mu(l,t)_{\pm}}\,dt^{2}+  dl^{2} +r(l)^{2}\,d\theta^{2} +
r(l)^{2}\sin^{2}(\theta)\,d\phi^{2} + a(t)\,d\Omega^{2}_{(D-4)},
\label{eq:proline}
\end{equation}
where $l(r)$ is the (signed) proper radial coordinate:
\begin{equation}
l(r)=\pm\int_{r_{0}}^{r}e^{\alpha(r\pme)/2}dr\pme. \label{eq:proprad}
\end{equation}
Here the lower sheet in figure \ref{fig:1} corresponds to the $-$
sign in (\ref{eq:proprad}) and the upper sheet corresponds to the
$+$, the throat located at $l=0$. Although continuous, the
profile curve is not manifest in this chart. It is desirable in
wormhole physics to retain the profile curve in calculations
since one can prescribe properties of the wormhole via
manipulation of the profile curve.

\begin{figure}[ht!]
\begin{center}
\includegraphics[bb=66 329 504 581, scale=0.8, keepaspectratio=true]{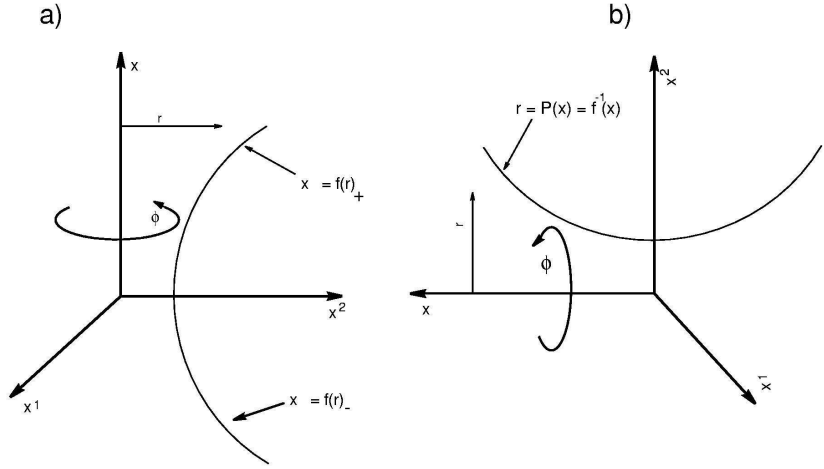}
\caption{{\small \textbf{a)} A cross section near the throat of a
wormhole embedding function for $t=\mbox{constant}$. The function
$x=f_{\pm}(r)$ represents the profile curve in the curvature
coordinates. \textbf{b)} By rotating this curve, we invert the
function $f(r)$ thus generating the profile curve $r=f^{-1}(x)=:P(x)$
in a coordinate system which is continuous across the throat.}}
\label{fig:2}
\end{center}
\end{figure}

\qquad We choose to employ another, ``global'', chart which we
briefly introduced in \cite{ref:debdas}. The relation between
this chart and the curvature coordinates of (\ref{eq:curvline})
is illustrated in figure \ref{fig:2}.
To generate this coordinate system, one starts with the standard
curvature coordinates. These consist of the upper and lower charts
with profile curves given by $f(r)_{+}$ for $x>0$ and $f(r)_{-}$ for
$x<0$ (refer to figure \ref{fig:2}a). Rotating the
$x\mbox{-}x^{2}$-plane about the $x^{1}$-axis of figure \ref{fig:2}a
by $\pi/2$, we can study the profile curve as parameterized by the
coordinate $x$. That is, the profile curve given by $r=P(x)$ (see
figure \ref{fig:2}b). The coordinate $x$ belongs to an interval
containing $x=0$, the throat of the wormhole. In this
parameterization, the function $P(x)$ is assumed to be continuous at
$x=0$ since an acceptable wormhole requires $\lim_{x\rightarrow
0^{+}}P(x)=\lim_{x\rightarrow 0^{-}}P(x)=P(0)$. Of course, the function
$P(x)$ is not completely arbitrary and must possess the correct
properties to describe a wormhole. These will be discussed below.
(Also, the profile function $P(x)$ must be at least of class $C^{4}$
to generate an acceptable metric.) We concentrate in this paper on the
region $x \geq 0$ as the physical results are the same for $x<0$ and
calculations can easily be modified to include $x<0$.

\qquad The surface of revolution generated by rotating the curve
in figure \ref{fig:2}b about the $x$-axis (the direction of
$\phi$) possesses the induced metric:
\begin{equation}
d\sigma^{2}=\ftr\,(dx)^{2}+ P(x)^{2}\,(d\phi)^{2}. \label{eq:twomet}
\end{equation}
\subsection{The general case} \label{subsec:gencase}
Any time dependent wormhole must possess a metric of the form
(\ref{eq:twomet}) on the $x\mbox{-}\phi$-sub-manifold. Therefore, the
corresponding spacetime metric may be written as
\begin{eqnarray}
ds^{2}&=&-e^{\gam}\,dt^{2}+e^{\lam}\left\{\left[1+P(x)_{,x}^{2}\right]\,dx^{2}
+P(x)^{2}\,d\theta^{2} +
P(x)^{2}\sin^{2}\theta\,d\phi^{2}\right\} \nonumber \\
&&+\at\,d\Omega_{(D-4)}^{2}\;. \label{eq:fullmet}
\end{eqnarray}
Notice that for constant $t$, (\ref{eq:fullmet}) possesses the
desired hypersurface metric as well as those conformally related
to it. At the wormhole throat, we must have $P(x)_{,x}=0$ and this
condition is easily satisfied in this coordinate system with well
behaved metric. (Note that in the standard curvature coordinates,
the throat condition $f(r)_{,r}\rightarrow \infty$ yields a
singular metric!) We denote the stellar boundary in the new chart
as $x=x_{b\pm}$ where the $\pm$ indicate the left and right
boundaries respectively. Note that $x_{b+}$ does not necessarily
have to equal $-x_{b-}\;\;$.

\qquad The Einstein field equations,
\begin{equation}
R^{\mu}_{\;\nu}-\frac{1}{2}R\,\delta^{\mu}_{\;\nu}+\Lambda\,\delta^{\mu}_{\;\nu}
=8\pi T^{\mu}_{\;\nu}, \label{eq:einsteqs}
\end{equation}
govern the geometry. These equations along with ansatz
(\ref{eq:fullmet}) yield the following differential equations:
\begin{eqnarray}
&&-\frac{1}{4}\left\{e^{-\gam}\left[\frac{1}{2}(D-5)(D-4)\left(\frac{\at_{,t}}{\at}\right)^{2}
+3(D-4)\lam_{,t}\frac{\at_{,t}}{\at} +3
\lam_{,t}^{2}\right]\right. \nonumber \\
&&+\frac{2(D-5)(D-4)}{\at}
+e^{-\lam}\left[1+P(x)^{2}_{,x}\right]^{-1}\left[\frac{4}{P(x)^{2}}-\lam_{,x}^{2}
-4\lam_{,x,x} \right. \nonumber \\
&& \left. -\frac{8}{P(x)}\lam_{,x}P(x)_{,x}\right] \left.
+4e^{-\lam}\left[1+P(x)^{2}_{,x}\right]^{-2}\p_{,x,x}\left[\p_{,x}\lam_{,x}
-\frac{2}{\p}\right]\right\} \nonumber \\
&&= \C T^{t}_{\; t}-\Lambda, \label{eq:fulleinst1}
\end{eqnarray}
\begin{eqnarray}
&&-\frac{1}{4}\left\{e^{-\gam}\left[\frac{1}{2}(D-7)(D-4)\left(\frac{\at_{,t}}{\at}\right)^{2}
+2(D-4)\lam_{,t}\frac{\at_{,t}}{\at}+3\lam_{,t}^{2}
\right.\right. \nonumber \\
&&\left.-2\gam_{,t}\lam_{,t}+2(D-4)\frac{\at_{,t,t}}{\at}+4\lam_{,t,t}
-(D-4)\gam_{,t}\frac{\at_{,t}}{\at} \right] \nonumber \\
&&+\frac{2(D-5)(D-4)}{\at}+e^{-\lam}\left[1+\p^{2}_{,x}\right]^{-1}
\left[\frac{4}{\p^{2}}-2\lam_{,x}\gam_{,x}-\lam_{,x}^{2} \right.
\nonumber \\
&&-\left.\left.\frac{4}{\p}\p_{,x}\left(\gam_{,x}+
\lam_{,x}\right)\right]\right\} =\C T^{x}_{\; x}-\Lambda,
\label{eq:fulleinst2}
\end{eqnarray}
\begin{equation}
-\frac{1}{4}e^{-\gam}\left[2\gam_{,x}\lam_{,t} -4\lam_{,x,t}
+(D-4)\gam_{,x}\frac{\at_{,t}}{\at}\right] = \C T^{t}_{\; x},
\label{eq:fulleinst3}
\end{equation}
\begin{eqnarray}
&&-\frac{1}{4}\left\{e^{-\gam}\left[
\frac{1}{2}(D-7)(D-4)\left(\frac{\at_{,t}}{\at}\right)^{2}
+2(D-4)\lam_{,t}\frac{\at_{,t}}{\at} +3\lam_{,t}^{2}
\right.\right. \nonumber \\
&&\left.+4\lam_{,t,t}+ 2(D-4)\frac{\at_{,t,t}}{\at}
-2\lam_{,t}\gam_{,t} -(D-4)\gam_{,t}\frac{\at_{,t}}{\at}\right]
\nonumber \\
&&+\frac{2(D-5)(D-4)}{\at} -2e^{-\lam}\left[1+\p^{2}_{,x}\right]^{-1}
\left[\lam_{,x,x}+\gam_{,x,x} \right. \nonumber \\
&&+\lam_{,x}\frac{\p_{,x}}{\p}
+\left.\frac{1}{2}\gam_{,x}^{2}+\frac{\p_{,x}}{\p}\gam_{,x}\right]
\nonumber \\
&&+\left. 2\p_{,x,x}e^{-\lam}\left[1+\p^{2}_{,x}\right]^{-2}
\left[\p_{,x}\left(\lam_{,x}+\gam_{,x}\right)-\frac{2}{\p}\right]
\right\} \nonumber \\
&&= \C T^{\theta}_{\;\theta} -\Lambda \equiv \C T^{\phi}_{\;\phi}
-\Lambda\;, \label{eq:fulleinst4}
\end{eqnarray}
\begin{eqnarray}
&&\frac{1}{4}\left\{e^{-\gam}\left[-\frac{1}{2}(D-8)(D-5)\left(
\frac{\at_{,t}}{\at}\right)^{2}+(D-5)\frac{\at_{,t}}{\at}
\left(\gam_{,t}-3\lam_{,t}\right)\right.\right. \nonumber \\
&&\left.-6\lam_{,t}^{2} +3\gam_{,t}\lam_{,t}-6\lam_{,t,t}-
2(D-5)\frac{\at_{,t,t}}{\at}\right] -\frac{2(D-5)(D-6)}{\at} \nonumber \\
&&+2e^{-\lam}\left[1+\p_{,x}^{2}\right]^{-1}
\left[4\lam_{,x}\frac{\p_{,x}}{\p}-\frac{2}{\p^{2}}
+\frac{1}{2}\left(\lam_{,x}^{2}+\gam_{,x}^{2}\right)\right.
\nonumber \\
&&\left.+2\lam_{,x,x} + \gam_{,x,x}+\frac{2}{\p}\gam_{,x}\p_{,x}
+\frac{1}{2}\gam_{,x}\lam_{,x}\right] \nonumber \\
&&\left. +2e^{-\lam}\left[1+\p^{2}_{,x}\right]^{-2}
\left[\frac{4}{\p}\p_{,x,x}
-\p_{,x}\p_{,x,x}\left(\gam_{,x}+2\lam_{,x}\right)\right]\right\}
\nonumber \\
&&=\C T^{A}_{\;A} -\Lambda. \label{eq:fulleinst5}
\end{eqnarray}
(Here the index $A$ is not summed!)

\qquad The conservation laws,
\begin{equation}
T^{\mu}_{\;\nu;\mu}= 0, \nonumber
\end{equation}
yield two non-trivial equations:
\begin{eqnarray}
T^{\mu}_{\;t;\mu}&=&\frac{1}{2} \left\{(D-4)\frac{\at_{,t}}{\at}
\left[T^{t}_{\;t}-T^{A}_{\;A}\right]
-\lam_{,t}\left[T^{x}_{\;x}+2T^{\theta}_{\;\theta}\right] \right.
\nonumber \\
&&+\frac{4}{\p}\p_{,x}T^{x}_{\;t} +3\lam_{,t}T^{t}_{\;t}
+3\lam_{,x}T^{x}_{\;t} +2\left[T^{x}_{\;t,x}+T^{t}_{\;t,t}\right]
\nonumber \\
&&+\ftr^{-1}\left[2\p_{,x}\p_{,x,x}T^{x}_{\;t}-\gam_{,x}e^{\gam-\lam}T^{t}_{\;x}
\right]\left.\right\} =0
\end{eqnarray}
and
\begin{eqnarray}
T^{\mu}_{\;x;\mu}&=&\frac{1}{2}\left\{2T^{t}_{\;x,t}+2T^{x}_{\;x,x}-
\lam_{,t}e^{\lam-\gam}\ftr
T^{x}_{\;t} \right. \nonumber \\
&&-\gam_{,x}T^{t}_{\;t}+T^{t}_{\;x}
\left[\gam_{,t}+2\lam_{,t}+(D-4)\frac{\at_{,t}}{\at}
\right] \nonumber \\
&&+T^{x}_{\;x}\left[\gam_{,x}+2\lam_{,x}+\frac{4}{\p}\p_{,x}\right]
-2T^{\theta}_{\;\theta}\left[\lam_{,x}+\frac{2}{\p}\p_{,x}\right]
\left. \right\} \nonumber \\ &=&0.
\end{eqnarray}

\qquad To study the singularity structure of the manifold, the
orthonormal Riemann components will be needed. These, as well as
those related by symmetry, are furnished by the following:
\begin{eqnarray}
R_{\hat{t}\hat{x}\hat{t}\hat{x}}&=&-\frac{1}{4}\left\{e^{-\gam}\left[
2\lam_{,t,t}+\lam_{,t}^{2}-\lam_{,t}\gam_{,t}\right]\right.\nonumber \\
&&+e^{-\lam}\left[1+\p^{2}_{,x}\right]^{-1} \left[\lam_{,x}\gam_{,x}
-\gam_{,x}^{2}-2\gam_{,x,x}\right]
\nonumber \\
&&\left.+2e^{-\lam}\left[1+\p^{2}_{,x}\right]^{-2}\gam_{,x}\p_{,x}\p_{,x,x}\right\},
\label{eq:rtxtx}
\end{eqnarray}
\begin{eqnarray}
R_{\hat{t}\hat{\theta}\hat{t}\hat{\theta}}&=&-\frac{1}{4}\left\{
e^{-\gam}\left[2 \lam_{,t,t}+ \lam_{,t}^{2}
-\gam_{,t}\lam_{,t}\right]\right. \nonumber \\
&&+\left. e^{-\lam}
\left[1+\p_{,x}^{2}\right]^{-1}\left[\lam_{,x}\gam_{,x}-
2\gam_{,x}\frac{\p_{,x}}{\p}\right]\right\}, \label{eq:rtthtth}
\end{eqnarray}
\begin{eqnarray}
R_{\hat{t}\hat{\theta}\hat{x}\hat{\theta}}&=&\frac{1}{2}e^{-\left[\gam+\lam\right]/2}
\left[1+\p^{2}_{,x}\right]^{-1/2}\left[\frac{1}{2}\lam_{,t}\gam_{,x}
-\lam_{,x,t}\right], \label{eq:rtthxth}
\end{eqnarray}
\begin{eqnarray}
R_{\hat{x}\hat{\theta}\hat{x}\hat{\theta}}&=&-\frac{1}{4\p}\left\{
2e^{-\lam}\left[1+\p^{2}_{,x}\right]^{-2}
\left[2\p_{,x,x}-\p\p_{,x}\p_{,x,x}\lam_{,x}\right] \right.
\nonumber \\
&& \left. +2e^{-\lam}\left[1+\p^{2}_{,x}\right]^{-1}
\left[\lam_{,x}\p_{,x} +\lam_{,x,x}\p\right] \right. \nonumber \\
&&\left. -e^{-\gam}\lam_{,t}^{2}\p\right\} , \label{eq:rxthxth}
\end{eqnarray}
\begin{eqnarray}
R_{\hat{\theta}\hat{\phi}\hat{\theta}\hat{\phi}}&=&\frac{1}{4\p^{2}}
\left\{e^{-\lam}\left[1+\p^{2}_{,x}\right]^{-1}
\left[4-\lam_{,x}^{2}\p^{2}-4\lam_{,x}\p_{,x}\p\right]\right.
\nonumber \\
&&\left. +e^{-\gam}\lam_{t}^{2}\p^{2} \right\},
\label{eq:rthphthph}
\end{eqnarray}
\begin{eqnarray}
R_{\hat{t}\hat{A}\hat{t}\hat{A}}&=&\frac{e^{-\gam}}{4\at^{2}}
\left[\gam_{,t}\at_{,t}\at -2\at_{,t,t}\at+\at_{,t}^{2}\right],
\label{eq:rtata}
\end{eqnarray}
\begin{eqnarray}
R_{\hat{t}\hat{A}\hat{x}\hat{A}}&=&
\frac{1}{4\at}e^{-\left[\gam+\lam\right]/2}
\left[1+\p^{2}_{,x}\right]^{-1/2}\gam_{,x}\at_{,t},
\label{eq:rtaxa}
\end{eqnarray}
\begin{eqnarray}
R_{\hat{x}\hat{A}\hat{x}\hat{A}}&=&\frac{1}{4}e^{-\gam}\lam_{,t}\frac{\at_{,t}}{\at},
\label{eq:rxaxa}
\end{eqnarray}
\begin{eqnarray}
R_{\hat{\theta}\hat{A}\hat{\theta}\hat{A}}&=&\frac{1}{4}e^{-\gam}\lam_{,t}
\frac{\at_{,t}}{\at}, \label{eq:rthatha}
\end{eqnarray}
\begin{eqnarray}
R_{\hat{A}\hat{B}\hat{A}\hat{B}}=&\frac{1}{4}\left[\frac{4}{\at}+
\left(\frac{\at_{,t}}{\at}\right)^{2}e^{-\gam}\right].
\label{eq:rabab}
\end{eqnarray}
Here hatted indices indicate quantities calculated in the
orthonormal frame. It is interesting to note that the Riemann
components are independent of the number of compact dimensions. A
singular manifold results if any of the following conditions
occur: $\at=0$, $\at_{,t} \rightarrow \pm\infty$,
$\at_{,t,t}\rightarrow \pm\infty$, $\gam\rightarrow -\infty$,
$\gam_{,t}\rightarrow \pm\infty$, $\gam_{,x}\rightarrow
\pm\infty$, $\gam_{,x,x}\rightarrow\pm\infty$, $\lam\rightarrow
-\infty$, , $\lam_{,t}\rightarrow\pm\infty$, ,
$\lam_{,x}\rightarrow\pm\infty$,
$\lam_{,t,t}\rightarrow\pm\infty$,
$\lam_{,x,x}\rightarrow\pm\infty$ and
$\lam_{,x,t}\rightarrow\pm\infty$.

\qquad Since the profile curve has not been specified, the above
system of equations may be used to model almost any spherically
symmetric spacetime with arbitrary number of compact dimension.
We wish here to study a wormhole system and therefore must, at
the very least, impose the condition that $x=0$ is a local
minimum of the profile curve. The sufficient conditions for such
a minimum are the following:
\begin{itemize}
\item $P(x)_{,x|_{x=0}}$ must vanish.
\item $P(x)_{,x}$, must change sign at the throat. $P(x)_{,x}>0$
 for $x>0$ and $P(x)_{,x}<0$ for $x<0$ (at least near the throat).
\item $P(x)$ must possess a positive second derivative at
least near the throat region.
\end{itemize}
Since there will be a cut-off at $x=x_{b\pm}$, where the solution
is joined to vacuum, no specification needs to be made for
$x>x_{b+}$ and $x<x_{b-}$.

\qquad With little loss of generality, we may model any near
throat geometry with the profile function
\begin{subequations}
\begin{align}
P(x)=&P_{0}+A^{2}x^{2n}\exp\left[h^{2}(x)\right],
\label{eq:prof} \\
P(x)_{,x}=&2A^{2}x^{2n}\exp\left[h^{2}(x)\right]
\left[\frac{n}{x}+h(x)h(x)_{,x}\right], \label{eq:dprof} \\
P(x)_{,x,x}=&2A^{2}x^{2n}\exp\left[h^{2}(x)\right]
\left\{2\left[\frac{n}{x}+h(x)\,h(x)_{,x}\right]^{2}\right.
\nonumber \\
&-\left.\frac{n}{x^2}+h(x)\,h(x)_{,x,x}+h(x)_{,x}^{2}\right\}.
\label{eq:ddprof}
\end{align}
\end{subequations}
Here $n=1,\,2,\,3,\,4, \ldots$, and $P_{0}>0$ is the radius of the
throat. The quantity $A$ is any non-zero constant and $h(x)$ a
sufficiently differentiable (at least $C^{4}$) arbitrary function. The
positive constant $n$ is restricted to integers for convenience. One
may also model wormholes for non-integer $n$ however, reasonable
physics places restrictions on the acceptable values of $n$ as will be
shown later. Such a function (\ref{eq:prof}) for $P(x)$ is capable of
describing infinitely many spherically symmetric wormholes at least
near the vicinity of the throat.

\qquad As mentioned in the introduction, an interesting property of
static wormholes is that such systems {\em must} violate the WEC even
if only in some arbitrarily small region. Time dependent analogues in
four dimensions have been examined in the context of inflationary
scenarios \cite{ref:kar2}, \cite{ref:roman} as well as general local
geometry at the near throat region \cite{ref:hocvispens}. However,
here we wish to study compact objects joined to an acceptable vacuum
with an arbitrary number of extra dimensions. Therefore, the
properties of the stress-energy tensor are mainly determined by
desirable physics and the junction conditions at the stellar boundary.
It turns out that the presence of this boundary imposes some
interesting restrictions on the wormhole structure.

\subsection{The quasi-static case} \label{subsec:quasistat}
\qquad The full field equations, given by (\ref{eq:fulleinst1} -
\ref{eq:fulleinst5}) are too formidable to solve in generality with
(\ref{eq:prof}) and reasonable stress-energy tensor. In four
dimensional General Relativity, the static assumption is often made to
keep the analysis tractable (for example, in many analytic studies of
stellar structure \cite{ref:weinberg} as well as wormholes
\cite{ref:visbook}). This assumption is motivated by the observational
fact that many stellar systems are slowly varying with time and
therefore the time-dependence may be ignored. Here we make a similar
assumption save for the fact that the radius of the extra dimensions
may vary with time (for examples of theories where extra dimensions
are explicitly static or time-dependent see \cite{ref:witten},
\cite{ref:shinshir}, \cite{ref:jgeddes}, \cite{ref:kan} and references
therein.) Put another way, since the solutions to the differential
equations are local, one may consider these solutions valid over a
time interval where the gravitational field does not evolve
appreciably in either the full space (the static case) or merely the
non-compact space (the quasi-static case). Below we consider both
situations. The latter case allows for the fact that, since the extra
dimensions are not on an equal footing with the non-compact ones, they
may evolve at a radically different rate than the non-compact
dimensions. (In a warped product manifold such as (\ref{eq:fullmet}),
it is the vacuum which will determine the behaviour of the
compactification radius, $a(t)$, if a vacuum-matter boundary is
present. It turns out that even if a vacuum appears static in the
non-compact $M^{4}$, it will support a compactification radius which
can vary rapidly with time as will be seen in section
\ref{subsubsec:tdepvac}.)

\qquad The above cases are of particular interest since static four
dimensional wormholes have now been well studied and many of their
properties determined. However, in light of the possibility of higher
dimensions, it is useful to extend studies to the higher dimensional
counter-parts of these cases. It turns out that the higher dimensional
static and ``quasi-static'' wormholes may differ considerably from the
four dimensional ones. Some results in this study are also valid for
trivial topology and therefore will shed some light on the behaviour
of ``regular'' static stars when one allows for the time dependent and
time independent extra dimensions.

\qquad The quasi-static configuration possesses line element:
\begin{equation}
ds^{2}=-e^{\gamma(x)}\,dt^{2}+\left[1+P(x)^{2}_{,x}\right]
\,dx^{2} + P(x)^{2}\,d\theta^{2} +
P(x)^{2}\sin^{2}\theta\,d\phi^{2} + \at\,d\Omega_{(D-4)}^{2}\;.
\label{eq:quasistatmet}
\end{equation}
and yield the corresponding Einstein equations:
\begin{subequations}
\begin{align}
&-\frac{1}{4}\left\{\frac{1}{2}e^{-\gamma(x)}(D-5)(D-4)\left(\frac{\at_{,t}}{\at}\right)^{2}
+\frac{2(D-5)(D-4)}{\at}+ \ftr^{-1}\frac{4}{\p^{2}} \right. \nonumber \\
&\left. -8\ftr^{-2}\frac{\p_{,x,x}}{\p} \right\}=8\pi
T^{t}_{\;t}-\Lambda, \label{eq:quasistat1} \\
&-\frac{1}{4}\left\{e^{-\gamma(x)}\left[\frac{1}{2}(D-7)(D-4)
\left(\frac{\at_{,t}}{\at}\right)^{2} + 2(D-4)\frac{\at_{,t,t}}{\at}
\right] +\frac{2(D-5)(D-4)}{\at}
\right. \nonumber \\
&+4\ftr^{-1}\left[\frac{1}{\p^{2}}-\gamma(x)_{,x}
\frac{\p_{,x}}{\p}\right]\left.\right\}=8\pi T^{x}_{\; x}
-\Lambda, \label{eq:quasistat2} \\
&-\frac{(D-4)}{4}e^{-\gamma(x)}\gamma_{,x}\frac{\at_{,t}}{\at} =
8\pi T^{t}_{\; x} \label{eq:quasistat3} \\
&-\frac{1}{4}\left\{e^{-\gamma(x)}\left[ \frac{1}{2} (D-7)(D-4)
\left(\frac{\at_{,t}}{\at}\right)^{2}
+2(D-4)\frac{\at_{,t,t}}{\at} \right] +\frac{2(D-5)(D-4)}{\at}
\right. \nonumber \\
&-2\ftr^{-1}\left[\gamma(x)_{,x,x}+\frac{1}{2}\gamma(x)^{2}_{,x}
+\frac{\p_{,x}}{\p}\gamma(x)_{,x}\right] \nonumber \\
&+2\p_{,x,x}\ftr^{-2}
\left[\p_{,x}\gamma(x)_{,x}-\frac{2}{\p}\right]\left.\right\} =
8\pi T^{\theta}_{\;\theta}-\Lambda\equiv 8\pi T^{\phi}_{\;\phi}
-\Lambda, \label{eq:quasistat4}  \\
&-\frac{1}{4}\left\{e^{-\gamma(x)} \left[\frac{1}{2}(D-8)(D-5)
\left(\frac{\at_{,t}}{\at}\right)^{2}
+2(D-5)\frac{\at_{,t,t}}{\at}\right] +\frac{2(D-5)(D-6)}{\at}
\right.
\nonumber \\
&+2\ftr^{-1}\left[\frac{2}{\p^{2}}-\frac{1}{2} \gamma(x)^{2}_{,x}
-\gamma(x)_{,x,x} -2\gamma(x)_{,x} \frac{\p_{,x}}{\p} \right]
\nonumber \\
&+2\ftr^{-2}\left[\gamma(x)_{,x}\p_{,x}\p_{,x,x}-
4\frac{\p_{,x,x}}{\p}\right]\left.\right\}=8\pi T^{A}_{\;A}-
\Lambda\: . \label{eq:quasistat5}
\end{align}
\end{subequations}
The conservation laws simplify to:
\begin{subequations}
\begin{align}
T^{\mu}_{\;t;\mu}=&\frac{1}{2}\left\{(D-4)\frac{\at_{,t}}{\at}
\left[T^{t}_{\;t}-T^{A}_{\;A}\right]+4\frac{\p_{,x}}{\p}T^{x}_{\;t}
+2\left[T^{x}_{\;t,x}+T^{t}_{\;t,t}\right] \right.
\nonumber \\
&\left.+\ftr^{-1} \left[2\p_{,x}\p_{,x,x}T^{x}_{\;t}
-\gamma(x)_{,x}e^{\gamma(x)}T^{t}_{\;x}\right]\right\}= 0,
\label{eq:quasistatcons1} \\
T^{\mu}_{\;x,\mu}=&\frac{1}{2}\left\{2T^{t}_{\;x,t}
+2T^{x}_{\;x,x} -\gamma(x)_{,x}T^{t}_{\;t}
+(D-4)\frac{\at_{,t}}{\at} T^{t}_{\;x}
\right. \nonumber \\
&\left.+\left[\gamma(x)_{,x}+4\frac{\p_{,x}}{\p}\right]T^{x}_{\;x}-
4T^{\theta}_{\;\theta}\frac{\p_{,x}}{\p}\right\}= 0,
\label{eq:quasistatcons2}
\end{align}
\end{subequations}
and the orthonormal Riemann components are furnished by:
\begin{subequations}
\begin{align}
R_{\hat{t}\hat{x}\hat{t}\hat{x}}=&\frac{1}{4}\left\{\ftr^{-1}
\left[\gamma(x)^{2}_{,x}+2\gamma(x)_{,x,x}\right] \right. \nonumber \\
&\left.+2\ftr^{-2}
\gamma(x)_{,x}\p_{,x}\p_{,x,x}\right\} \label{eq:quasistatrtxtx} \\
R_{\hat{t}\hat{\theta}
\hat{t}\hat{\theta}}=&\frac{1}{2}\ftr^{-1}\gamma(x)_{,x}
\frac{\p_{,x}}{\p} \label{eq:quasistatrtthtth} \\
R_{\hat{x}\hat{\theta}
\hat{x}\hat{\theta}}=&-\ftr^{-2}\frac{\p_{,x,x}}{\p^{2}}
\label{eq:quasistatrxthxth} \\
R_{\hat{\theta}\hat{\phi}\hat{\theta}\hat{\phi}}=&
\frac{1}{\p^{2}}\ftr^{-1}
\label{eq:quasistatrthphthph} \\
R_{\hat{t}\hat{A}\hat{t}\hat{A}}=&\frac{e^{-\gamma(x)}}{4\at^{2}}
\left[\at^{2}_{,t}
-2\at\at_{,t,t}\right] \label{eq:quasistatrtata} \\
R_{\hat{t}\hat{A}\hat{x}\hat{A}}=&\frac{1}{4\at^{2}}
e^{-\gamma(x)/2}\ftr^{-1/2}
\gamma(x)_{,x}\at_{,t} \label{eq:quasistatrtaxa} \\
R_{\hat{A}\hat{B}\hat{A}\hat{B}}=&\frac{1}{4}\left[\frac{4}{\at}+
\left(\frac{\at_{,t}}{\at}\right)^{2}e^{-\gamma(x)}\right].
\label{eq:quasistatrabab}
\end{align}
\end{subequations}

By analyzing the orthonormal Riemann components
(\ref{eq:quasistatrtxtx}- \ref{eq:quasistatrthphthph}) corresponding
to the four non-compact dimensions, it can be seen that this indeed
describes a system which is static in the first four dimensions. From
equations (\ref{eq:quasistatrtata}-\ref{eq:quasistatrabab}) it is
clear that time dependence of the gravitational field is experienced
only by parallel transport of vectors around closed loops which
venture into the higher dimensions. Therefore, an observer unaware of
the presence of the extra dimensions will ``see'' only a static
wormhole or star.

\section{Physical structure} \label{sec:physstruct}

\subsection{The vacuum} \label{subsec:vacstruct}
Isolated systems such as those described here should possess some
boundary differentiating the vacuum from the matter which
supports the wormhole. In four dimensions there is the Birkhoff's
theorem which dictates that the only acceptable spherically
symmetric vacuum is locally equivalent to the Schwarzschild (or
Schwarschild-(anti) deSitter for non-zero cosmological constant)
vacuum \cite{ref:birk}. It turns out that a similar theorem holds
for any higher number of {\em non-compact} dimensions
\cite{ref:bronik}, \cite{ref:schmidtbirk}, \cite{ref:dasdeb}.

\qquad By studying the vacuum version of equations
(\ref{eq:quasistat1})-(\ref{eq:quasistat5}) we encounter the
possibility for several vacua which fall into two general classes.
From equation (\ref{eq:quasistat3}) we see that for $D>4$,
$G^{t}_{\;x}=0$ implies that either $\gamma\equiv\gamma_{0}$ or
that $a(t)\equiv \varepsilon$, where $\gamma_{0}$ and
$\varepsilon$ are constants. For both cases, we begin by solving
the linear combination
\begin{equation}
G^{t}_{\;t}-G^{x}_{\;x}=0\;. \label{eq:lincomb}
\end{equation}

\subsubsection{The vacuum $\gamma\equiv\gamma_{0}$}
\label{subsubsec:tdepvac} \qquad In this situation the equation
(\ref{eq:lincomb}) provides the following set of equations
\begin{subequations}
\begin{align}
&\left[\frac{\at_{,t}}{\at}\right]^{2}-2\frac{\at_{,t,t}}{\at}=
k\frac{e^{\gamma_{0}}}{D-4}=:k_{0}, \label{eq:aeq} \\
&8\frac{\p_{,x,x}}{\p}\ftr^{-2}=k. \label{eq:peq}
\end{align}
\end{subequations}
At the moment, the separation constant, $k$, may be positive, zero, or
negative corresponding respectively to elliptic, parabolic and
hyperbolic behaviour of the compact dimensions.

\qquad It is found that one cannot satisfy all vacuum field
equations for the cases $k>0$ and $k=0$ and therefore we
concentrate only on the $k<0$ scenario. The solutions of
(\ref{eq:aeq}) and (\ref{eq:peq}) for $k<0$ are:
\begin{equation}
a(t)=\left[a_{0}e^{\omega t} + b_{0}e^{-\omega t}\right]^{2}
\label{eq:vaca1}
\end{equation}
and
\begin{equation}
\p=\left[\frac{8}{|k|}-\left(x-x_{0}\right)^{2}\right]^{1/2},
\label{eq:vacp1}
\end{equation}
with $\omega:=\sqrt{\frac{|k|e^{\gamma_{0}}}{4(D-4)}}$ and $a_{0}$
and $b_{0}$ constants of integration. After long calculations, all
vacuum equations may be solved provided that:
\begin{equation}
|k|=\frac{8}{D-2}\,\Lambda > 0\;\;\;\;\mbox{and}\;\;\;\;
a_{0}b_{0}=\frac{D-4}{|k|} > 0\nonumber
\end{equation}
indicating a deSitter $R$-type vacuum with
$\frac{D-2}{\Lambda}>(x-x_{0})^{2}\geq 0$. The constant $x_{0}$
corresponds to the ``centre'' of the deSitter shell. The surface
$(x-x_{0})^{2}=\frac{D-2}{\Lambda}$ is not an event horizon since, in
the coordinate system of (\ref{eq:fullmet}) the presence an event
horizon is indicated by the condition $g_{tt}=0$. The vacuum metric,
(\ref{eq:quasistatmet}), is therefore given by
\begin{eqnarray}
ds^{2}_{vac_1}&=&-e^{\gamma_{0}}\,dt^{2}+
\left[1+\frac{(x-x_{0})^{2}}{\frac{D-2}{\Lambda}-(x-x_{0})^{2}}\right]\,dx^{2}
+ \left[\frac{D-2}{\Lambda}-(x-x_{0})^2\right]\,d\Omega_{(2)}^{2}
\nonumber \\
&&+\left[a_{0}e^{\omega
t}+\frac{(D-4)(D-2)}{8\Lambda\,a_{0}}e^{-\omega
t}\right]^{2}\,d\Omega_{(D-4)}^{2}. \label{eq:tdepvac}
\end{eqnarray}
This vacuum differs from the vacua considered below and therefore
there is no unique vacuum when one allows for the possibility of
compact extra dimensions.

\begin{figure}[ht!]
\begin{center}
\includegraphics[bb=100 447 453 635, clip, scale=0.8, keepaspectratio=true]{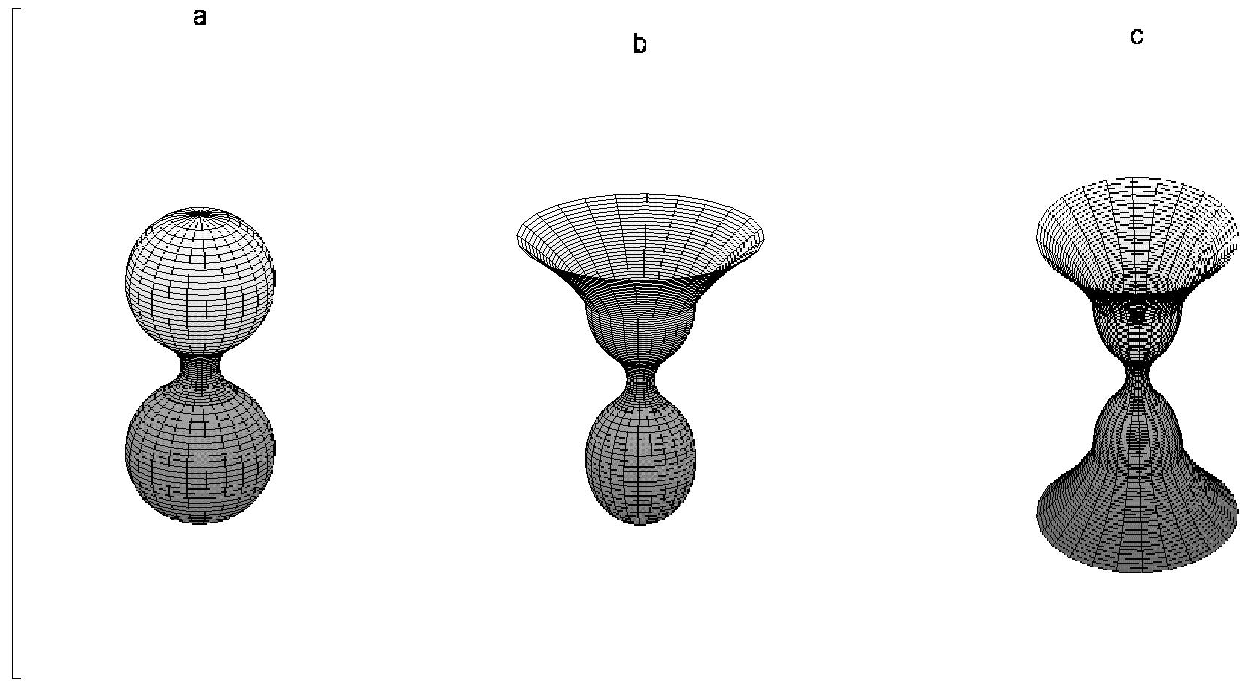}
\caption{{\small Rotated profile curves representing: \textbf{a)} A
dumbbell universe consisting of two spatially-closed universes
connected by a narrow throat, \textbf{b)} A spatially-closed universe
connected to an open universe, \textbf{c)}} Two open universes
connected by a wormhole.} \label{fig:3}
\end{center}
\end{figure}

\qquad If this is to be considered the global vacuum, then the
wormhole system must actually describe a ``dumbbell'' wormhole
\cite{ref:vishoc}, \cite{ref:hocvis}. Although topologically trivial,
such systems are considered wormholes in the literature
\cite{ref:hocvispens}. This scenario is shown qualitatively in figure
\ref{fig:3}a. Alternately, one can patch one or both spatially-closed
universes to an open (non-vacuum) universe thus creating the systems
shown in figure \ref{fig:3}b and \ref{fig:3}c.

\subsubsection{The vacuum $a=\varepsilon$}
\qquad For this case, the equation (\ref{eq:lincomb}) provides a
useful relation between $\gamma(x)_{,x}$ and the derivatives of $\p$:
\begin{equation}
\gamma(x)_{,x}=2\frac{\p_{,x,x}}{\p_{,x}\ftr}\;. \label{eq:gampreln}
\end{equation}
This may be used in the equation $G^{\theta}_{\;\theta}-G^{A}_{\;A}=0$
yielding a differential equation for $\p$ only. The resulting
autonomous equation may be solved implicitly for $\p$:
\begin{equation}
\pm\int\left[\frac{\frac{D-5}{3\varepsilon}P^{3}
-E_{0}}{-\frac{D-5}{3\varepsilon}P^{3}+P+E_{0}}\right]^{1/2}dP
=x-x_{0}\;, \label{eq:pvac2}
\end{equation}
with $E_{0}$ a constant resulting from integration. We consider
first the $D=5$ geometry.

\qquad For the case $D=5$ one must demand that $E_{0}<0$ and
(\ref{eq:pvac2}) may be readily integrated. The profile curve for this
case is
\begin{equation}
P(x)_{D=5}= \left[|E_{0}|+\frac{(x-x_{0})^2}{4|E_{0}|}\right],
\label{eq:p5dstat}
\end{equation}
which, with (\ref{eq:gampreln}) gives the spatially-open vacuum
with \emph{zero} cosmological constant:
\begin{eqnarray}
ds^{2}_{5D\;vac}&=&-\frac{(x-x_{0})^{2}}{4|E_{0}|^{2}
+(x-x_{0})^{2}}\, dt^{2}
+\left[1+\frac{(x-x_{0})^{2}}{4|E_{0}|^{2}}
\right]\,dx^{2} \nonumber \\
&&+\left[|E_{0}|+\frac{(x-x_{0})^{2}}{4|E_{0}|}
\right]^{2}\,d\Omega_{(2)}^{2}+ \varepsilon\, d\psi_{(0)}^{2}.
\label{eq:5dstatmet}
\end{eqnarray}
This is essentially the Schwarzschild solution crossed with one
compact dimension and $|E_{0}|$ is related to the gravitating mass,
$M$, via $|E_{0}|:=2M$. The event horizon exists at $x=x_{0}$.

\qquad For the case $E_{0}=0$ and $D>5$ the equation
(\ref{eq:pvac2}) may also be explicitly solved for $\p$ as
\begin{equation}
\p=\left[\frac{3\varepsilon}{D-5}-\left(x-x_{0}\right)^{2}
\right]^{1/2}. \label{eq:pexplic}
\end{equation}
This, along with equation (\ref{eq:gampreln}) yields the metric
function:
\begin{equation}
e^{\gamma(x)}=\kappa_{0}^{2}\left(x-x_{0}\right)^{2},\;\;\;\;\;\;
\kappa_{0} \mbox{ a constant} \label{eq:metfunvac2}
\end{equation}
and therefore the vacuum line element, (\ref{eq:quasistatmet}), is
given by
\begin{eqnarray}
ds^{2}_{D>5\; vac}&=&-\kappa_{0}^{2}\left(x-x_{0}\right)^{2}\,dt^{2} +
\left[1+\frac{(x-x_{0})^{2}}{\frac{3\varepsilon}{D-5}-
(x-x_{0})^{2}}\right]\,dx^{2} \nonumber \\
&& + \left[\frac{3\varepsilon}{D-5}- (x-x_{0})^{2}
\right]\,d\Omega^{2}_{(2)}+\varepsilon\,d\Omega^{2}_{(D-4)}.
\label{eq:staticvac}
\end{eqnarray}
This metric solves the vacuum field equations with cosmological
constant:
\begin{equation}
\Lambda=\frac{(D-2)(D-5)}{2\varepsilon}\;. \label{eq:statlambda}
\end{equation}
The vacuum is spatially-closed with $\frac{3(D-2)}{2\Lambda}  >
(x-x_{0})^{2} \geq 0$. The surface $x=x_{0}$ is a cosmological horizon
as is found in the deSitter universe and is of no concern if
considering a traversable wormhole.

\qquad Although (\ref{eq:pvac2}) cannot be explicitly integrated
for the general case, it can be shown that the metrics described
by this equation are similar to the four dimensional Kottler
\cite{ref:kottler} solution with the value of the cosmological
constant dictated by (\ref{eq:statlambda}). The relation to the
Kottler solution is not unexpected since, for constant $a(t)$,
one has a direct product manifold. For $D=5$ the manifold
structure is $M^{4}\times S^{1}$ with $S^{1}$ being intrinsically
flat, thereby yielding zero cosmological constant as above.
Similarly, for $D>5$, the compact dimensions yield an $S^{D-4}$
manifold, demanding positive cosmological constant.

\qquad An acceptable interval for (\ref{eq:pvac2}) with $D > 5$
and $E_{0} \neq 0$ is provided by the reality condition on
(\ref{eq:pvac2}). Note that this restriction again implies a
spatially-closed universe.

\subsection{Wormhole structure}
\qquad In this section we discuss the matter region given by $x_{b}
\geq x \geq 0$. We attempt to limit, or eliminate where possible,
violations of the WEC as well as demonstrate how singularities and
black-hole-event horizons may be avoided.

\qquad At the boundary, $x=x_{b}$, we employ the junction
conditions of Synge \cite{ref:syngebook} which read:
\begin{equation}
T^{\mu}_{\;\nu}\:\hat{n}^{\nu}_{\;\:|_{x=x_{b}}}=0,
\label{eq:syngcond}
\end{equation}
where $\hat{n}^{\nu}$ is an outward pointing unit normal to the
boundary. Explicitly, these conditions yield the following:
\begin{subequations}
\begin{align}
T^{t}_{\;x|_{x=x_{b}}}=0, \label{eq:syngcond1} \\
T^{x}_{\;x|_{x=x_{b}}}=0. \label{eq:syngcond2}
\end{align}
\end{subequations}
The junction condition (\ref{eq:syngcond1}) dictates (referring
to (\ref{eq:quasistat3})) that either $\gamma(x)_{,x}=0$ at the
stellar boundary or that $a(t)\equiv \varepsilon$. We consider
both cases. The first will be patched to the vacuum
(\ref{eq:tdepvac}) and the latter to one of the static vacua.

\qquad We wish the stellar material to satisfy the weak energy
condition (WEC) which is given by
\begin{equation}
T_{\mu\nu}V^{\mu}V^{\nu} \geq 0\;\;\;\; \forall \;\;\;\;
\mbox{time-like}\;\; V^{\alpha}. \label{eq:WEC}
\end{equation}
In the case of a non-diagonal stress-energy tensor,
(\ref{eq:WEC}) yields the conditions:
\begin{equation}
T^{i}_{\;i} - T^{t}_{\; t}\geq 0 \;\;\;\; \mbox{and} \;\;\;\;
T^{t}_{\;t} \leq 0 \, , \;\;\;\; i\neq x\:, \;\;\; \mbox{no
summation}, \label{eq:econd1}
\end{equation}
along with one of the following:
\begin{subequations}
\begin{align}
&\mbox{for } T^{x}_{\;x} \leq 0: \;\;\;T^{x}_{\;x}- T^{t}_{\;t}
\geq 2\frac{e^{\gamma(x)/2}}{\ftr^{1/2}}|T^{t}_{\; x}|,
\label{eq:econd2a} \\
&\mbox{for } T^{x}_{\;x} > 0 \mbox{ either }:  \nonumber \\
&\frac{e^{\gamma(x)}}{\ftr} (T^{t}_{\;x})^{2} +T^{x}_{\;x} T^{t}_{\;t}
\leq 0
\label{eq:econd2b} \\
\mbox{or} \nonumber \\
&0 <\left[\frac{e^{\gamma(x)}}{\ftr}
\left(T^{t}_{\;x}\right)^{2}+T^{x}_{\;x}T^{t}_{\;t}\right]^{1/2} \leq
\frac{e^{\gamma(x)/2}}{\ftr^{1/2}}|T^{t}_{\; x}| - T^{x}_{\;x}.
\label{eq:econd2c}
\end{align}
\end{subequations}
One could also write the above in the orthonormal frame which
would eliminate the factors of the metric. However, for the
analysis here it is more convenient to use the above form.

\subsubsection{The $a\equiv\varepsilon$ wormhole}
\qquad This case corresponds to a purely static wormhole. However, it
will be shown that the presence of compact extra dimensions yield
restrictions not present in the four dimensional static wormhole.

\qquad We attempt to satisfy the WEC (\ref{eq:WEC}) in the
vicinity of the throat. If WEC violation occurs away from the
throat, at $|x|=|x_{*}|$ say, one may attempt to place the
stellar boundary at some $|x|<|x_{*}|$ to eliminate the WEC
violation. An expansion near the throat utilizing
(\ref{eq:quasistat1}) and (\ref{eq:prof} yields:
\begin{eqnarray}
-8\pi T^{t}_{\;t}&\approx& \frac{1}{P^{2}_{0}}-\frac{2}{D-2}
\Lambda - \frac{4}{P_{0}}\left[2n^{2}-n\right]A^{2}e^{h^{2}(0)}
x^{2n-2} \nonumber \\
&&-\frac{8}{P_{0}}\left[2n^2+n\right]A^{2}e^{h^{2}(0)}h(0)h(x)_{,x\,|x=0}
x^{2n-1}+\mathcal{O}(x^{2n}), \label{eq:statdens}
\end{eqnarray}
where the extra dimensions affect the matter energy density via
the cosmological constant term (from equation
\ref{eq:statlambda}). Here we see why reasonable physics restricts
acceptable values of $n$ if one were to consider arbitrary real
numbers for $n$. For $n<1$ a singular density (as well as
$P(x)_{,x,x}$) results at the wormhole throat which yields a
singular manifold. However, physical results in this paper are
unchanged by dropping the integer restriction on $n$.

\qquad The best scenario from the point of view of energy
conditions is that where $n>1$. This gives the largest possible
energy density at the throat. Exactly at the throat ($x=0$)
(\ref{eq:statdens}), with $n>1$, gives:
\begin{equation}
\left[-8\pi T^{t}_{\;t}\right]_{|x=0}=\frac{1}{P_{0}^{2}}
-\frac{2}{D-2}\Lambda . \label{eq:statdensnrthrt}
\end{equation}
Demanding that (\ref{eq:statdensnrthrt}) be non-negative (as
required by WEC) and noting the relation between $\Lambda$ and
$\varepsilon$ as dictated by (\ref{eq:statlambda}) yields a
particularly interesting result:
\begin{equation}
P_{0}^{2}< \frac{\varepsilon}{D-5}\;. \label{eq:throatlimit}
\end{equation}
That is, if the energy density as seen by static observers is to
be positive for $D>5$, \emph{the radius of the wormhole throat
must be less than the radius of the extra dimensions}. For $D=5$
and $n>1$, any radius is allowed. The greater the number of extra
dimensions, the smaller a WEC respecting throat must be. If one
considers $n=1$ then the radius must be even smaller than that
dictated by (\ref{eq:throatlimit}) since the third term in
(\ref{eq:statdens}) contributes. Note that the above restrictions
are \emph{stronger} than the purely geometric restriction in which
the radius of the wormhole throat must be less than the radius of
the closed universe (which in this case would read $P_{0}^{2} <
3\varepsilon/(D-5)$.)

\qquad Next, we concentrate on the parallel pressure, $T^{x}_{\;x}$.
This function will be derived in such a way as to satisfy as many
physical requirements as possible (for example, no singularities). As
well, junction condition (\ref{eq:syngcond2}) must be met.

\qquad A necessary condition for the absence of singularities is
that the derivative, $\gamma(x)_{,x}$, be finite. From the
equation (\ref{eq:quasistat2}) we find
\begin{equation}
\gamma(x)_{,x}=\frac{\p}{\p_{,x}}\left\{\ftr\left[\taux\right]+\frac{1}{\p^{2}}\right\},
\label{eq:gamcomx}
\end{equation}
where we have used the relation (\ref{eq:statlambda}) to
eliminate $\varepsilon$. For (\ref{eq:gamcomx}) to be finite at
the throat, we must demand that the term in braces vanish at
least as fast as $\p_{,x}$ near the throat. The behaviour of
$\p_{,x}$ near the throat is of order $\mathcal{O}(x^{2n-1})$ and
therefore we must demand that $\taux\approx -\frac{1}{P_{0}^{2}}
+\mathcal{O}(x^{2n-1})$. Also, $T^{x}_{\;x}$ must vanish at
$x=x_{b}$. In terms of arbitrary functions, these condition may be
enforced via
\begin{eqnarray}
\taux&=&-\frac{1}{\p^{2}}\cos\left[\left(\frac{x}{x_{b}+\alpha}\right)^{(2n)p}
\frac{m\pi}{2}\right] \cosh\left[x^{s}\xi(x)\right] \nonumber \\
&&+\kappa x^{q}\left(x-x_{b}\right)^{w}e^{\zeta(x)}.
\label{eq:tauofx}
\end{eqnarray}
Here, $\xi(x)$ and $\zeta(x)$ are arbitrary, differentiable functions.
The constants are restricted as follows: $p\geq 1/2$, $m$ an odd
integer, $s\geq n$, $q\geq 2n$, $w$ a positive integer and $\kappa$ a
constant. The constant, $\alpha$, is set by the requirement that the
right-hand-side of (\ref{eq:tauofx}) is equal to $-2/(D-2)\;\Lambda$
when $x=x_{b}$. The second term on the right-hand-side of
(\ref{eq:tauofx}) is not required but is useful as a ``fine tuning''
function if we wish the solution to obey the WEC throughout most of
the stellar bulk (if the first term is negative with too large a
magnitude to respect the WEC, one may make this second term large and
positive).

\begin{figure}[ht!]
\begin{center}
\includegraphics[bb=90 116 445 630, clip, scale=0.6, keepaspectratio=true]{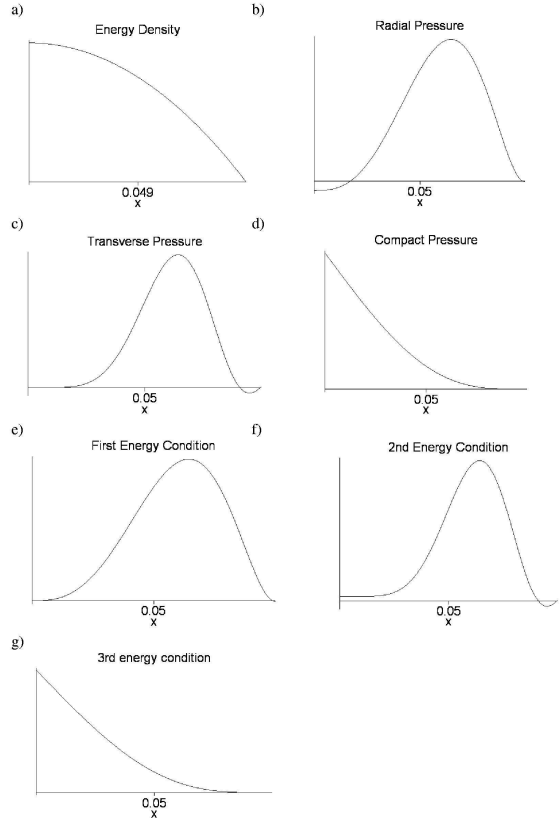}
\caption{{\small An example of a static wormhole. The graphs
indicate: a) $-T^{t}_{\;t}$, b) $T^{x}_{\;x}$, c)
$T^{\theta}_{\;\theta}$, d) $T^{A}_{\;A}$, e)
$T^{x}_{\;x}-T^{t}_{\;t}$, f)
$T^{\theta}_{\;\theta}-T^{t}_{\;t}$, g)$T^{A}_{\;A}-T^{t}_{\;t}$.
The functions are plotted from $x=0$ to $x=x_{b}$ (horizontal
axis). The vertical scales differ for each plot. These graphs
were produced for $D=7$. }} \label{fig:stateconds}
\end{center}
\end{figure}

\quad To study the WEC near the throat, $T^{x}_{\;x}$ is expanded
in powers of $x$:
\begin{equation}
8\pi T^{x}_{\;x}\approx
-\frac{1}{P_{0}^2}+\frac{2}{D-2}\Lambda-\frac{\xi(0)^2}{2P_{0}^2}x^{2s}
+ \kappa x_{b}^{w}e^{\zeta(0)} x^{q} - \frac{2}{P_{0}^{3}}A^{2}
e^{h^{2}(0)} x^{2n} +\mathcal{O}(x^{2n+1}). \label{eq:statpres}
\end{equation}
Equations (\ref{eq:statdens}) and (\ref{eq:statpres}) furnish the
WEC:
\begin{equation}
8\pi\left(T^{x}_{\;x}-T^{t}_{\;t}\right) \approx
-\frac{4}{P_{0}}\left[2n^2-n\right]A^{2}e^{h^{2}(0)} x^{2n-2} +
\mbox{higher order in }x.
\end{equation}
Therefore, although the WEC \emph{is} respected at the throat
(for $n>1$), slightly away from the throat, there must be a region
of WEC violation. This violation regional may be made arbitrarily
small (but non-zero). This is similar to results for the four
dimensional static case \cite{ref:kuhfittig}, \cite{ref:debdas}.
It can be shown that the models discussed in this section can
satisfy the WEC throughout most of the stellar bulk and we do so
by the specific example in the figure \ref{fig:stateconds}. One
may also patch to different WEC respecting layers (depending on
the physical stellar model one wishes to study), again yielding a
WEC respecting solution throughout most of the bulk.

\qquad In figure \ref{fig:stateconds}, we plot the eigenvalues of the
stress-energy tensor as well as the corresponding energy conditions.
Energy condition violation near the throat is minimal and does not
show up in figure \ref{fig:stateconds}e. There is also minor WEC
violation near the stellar boundary as is evident from figure
\ref{fig:stateconds}f. This may be eliminated by patching the solution
to an intermediate layer of WEC respecting matter and then patching
this layer to the vacuum. The transverse pressure,
$T^{\theta}_{\;\theta}$, and compact pressure, $T^{A}_{\;A}$ are
defined via equations (\ref{eq:quasistatcons2}) and
(\ref{eq:quasistat5}) respectively.

\qquad At the stellar boundary, the junction conditions of Synge
are met and the metric must match the static vacuum. The metric
function, $\gamma(x)$, at the boundary is given by the integral
of (\ref{eq:gamcomx}):
\begin{equation}
\gamma(x_{b})=\int_{x^{\prime}=0}^{x_{b}}
\frac{P(x^{\prime})}{P(x^{\prime})_{,x^{\prime}}}
\left\{\left[1+P(x^{\prime})^{2}_{,x^{\prime}}\right]
\left[\tauxp\right]+\frac{1}{P(x^{\prime})^{2}}\right\}dx^{\prime}
+g_{0}, \label{eq:gambound}
\end{equation}
with $g_{0}$ a constant. The metric at the boundary,
(\ref{eq:gambound}), may be joined to the vacuum of
(\ref{eq:staticvac}) via the identification:
\begin{equation}
\gamma(x_{b}):=2\ln\left[|\kappa_{0}(x_{b}-x_{0})|\right] \;\;\;
\mbox{or }\;\;\;
|\kappa_{0}|:=\frac{e^{\gamma(x_{b})/2}}{|x_{b}-x_{0}|}
\end{equation}
which sets the value of $\kappa_{0}$. This essentially corresponds to
a re-scaling of the time coordinate. A similar identification may be
utilised to join to the five dimensional vacuum of
(\ref{eq:5dstatmet}).

\qquad For continuity of other metric components at the boundary, it
is required that $P(x)$ and $P(x)_{,x}$ be continuous at $x_{b}$. This
may most easily be accomplished by fixing one of the many free
parameters of the profile curve. The continuity of $\p$ yields the
condition:
\begin{equation}
\left[\frac{3(D-2)}{2\Lambda}-(x_{b}-x_{0})^2\right]^{1/2}=P_{0}+
A^{2}x_{b}^{2n}e^{h^{2}(x_{b})}.
\end{equation}
Which can easily be satisfied by setting $A$, for example:
\begin{equation}
A^{2}=\frac{\left[\frac{3(D-2)}{2\Lambda}-(x_{b}-x_{0})^2\right]^{1/2}-
P_{0}}{x_{b}^{2n}}e^{-h^{2}(x_{b})}. \label{eq:Aboundary}
\end{equation}
The continuity of $\p_{,x}$ requires that
\begin{equation}
\frac{-(x_{b}-x_{0})}{\left[\frac{3(D-2)}{2\Lambda}-(x_{b}-x_{0})^2\right]^{1/2}}
=2A^{2}x_{b}^{2n}e^{h^{2}(x_{b})}
\left[h(x_{b})h(x)_{,x|x=x_{b}}+\frac{n}{x_{b}}\right],
\label{eq:contpx}
\end{equation}
which may be satisfied by setting any one of the remaining free
parameters ($x_{0}$, $x_{b}$, $h(x_{b})$ or $\Lambda$). For example,
using (\ref{eq:Aboundary}) in (\ref{eq:contpx}) one could place a
boundary condition on $h(x)$:
\begin{equation}
h(x_{b})h(x)_{,x|x=x_{b}}=
\frac{-(x_{b}-x_{0})}{\left[\frac{3(D-2)}{2\Lambda}-(x_{b}-x_{0})^2\right]^{1/2}
\left\{\left[\frac{3(D-2)}{2\Lambda}-(x_{b}-x_{0})^2\right]^{1/2}-P_{0}
\right\}} -\frac{n}{x_{b}}. \label{eq:boundaryh}
\end{equation}
A simple way to enforce this condition (although by no means the only
way) is to set $h(x):=\sqrt{2xF_{0}+ (x-x_{b})^{2p}u(x)}$ with $F_{0}$
given by the right-hand-side of (\ref{eq:boundaryh}), $u(x)$ an
arbitrary differentiable function and $p$ a sufficiently large
constant.

\subsubsection{The time-dependent wormhole}

\qquad We consider here the quasi-static wormhole with the scale
factor dictated by (\ref{eq:vaca1}). Again we wish to study the
WEC and begin by expanding $T^{t}_{\;t}$ near the wormhole throat:
\begin{eqnarray}
-8\pi T^{t}_{\;t}&\approx& \frac{1}{P^{2}_{0}}- \Lambda -
\frac{4}{P_{0}}\left[2n^{2}-n\right]A^{2}e^{h^{2}(0)}
x^{2n-2} \nonumber \\
&&-\frac{8}{P_{0}}\left[2n^2+n\right]A^{2}e^{h^{2}(0)}h(0)h(x)_{,x\,|x=0}
x^{2n-1} \nonumber \\
&&+
\frac{1}{4}\left\{\frac{1}{2}e^{-\gamma(0)}\left[1-\gamma(x)_{,x|x=0}x\right]
(D-5)(D-4) \left[\frac{\at_{,t}}{\at}\right]^{2}
+\frac{2(D-5)(D-4)}{\at}
\right\} \nonumber \\
&&+ \mathcal{O}(x^{2n}). \label{eq:tdepTtt}
\end{eqnarray}
Exactly at the throat, this quantity yields:
\begin{eqnarray}
-8\pi T^{t}_{\;t|x=0}&=&\frac{1}{P_{0}^{2}}-\Lambda
+\frac{1}{8}e^{-\gamma(0)}(D-5)(D-4)\left[\frac{\at_{,t}}{\at}\right]^{2}
+\frac{(D-5)(D-4)}{2\at} \nonumber \\
&&- \frac{4}{P_{0}}\left[2n^{2}-n\right]A^{2}e^{h^{2}(0)}\,
\delta_{n1}. \label{eq:tdepTttthroat}
\end{eqnarray}
Notice that, although the $\Lambda$ term makes a negative
contribution to the energy density of the matter ($\Lambda$ must
be positive as dictated by the vacuum), all other terms make
positive contributions (for $n>1$) and therefore conditions are
quite favourable to have a WEC respecting wormhole. It should be
noted that, for $D=5$, the scale factor makes no contribution to
$T^{t}_{\;t}$. However, unlike the static case, for $D>5$ we have
new time-dependent term which \emph{seemingly} may be made
arbitrarily large and makes a positive contribution to the energy
density. We consider below only the cases for $n>1$ since it is
easy to check that any energy condition violation is more severe
for $n=1$.

\qquad To satisfy the boundary condition (\ref{eq:syngcond1}), we
write:
\begin{equation}
\gamma(x)_{,x}=(x_{b}-x)^{l}g(x), \label{eq:tdepgx}
\end{equation}
with $l\geq 2$ or $l=1$ and $g(x)$ differentiable but otherwise
arbitrary which helps in eliminating singularities. Also, from
the equation (\ref{eq:quasistat2}) we find that
\begin{equation}
8\pi T^{x}_{\;x|x=x_{b}}=\frac{1}{D-2}\Lambda -
\left[1+P(x)^{2}_{,x|x=x_{b}}\right]^{-1}\frac{1}{P(x_{b})^{2}}.
\label{eq:tdeptxxbound}
\end{equation}
Here we have used (\ref{eq:tdepgx}) and the fact that
$\gamma(x_{b})\equiv \gamma_{0}$ as required for continuity of the
metric at the boundary. From the solution of the vacuum equations, the
time-dependent terms in $T^{x}_{\;x}$ give a constant, related to the
cosmological constant, at $x=x_{b}$ (or equivalently when
$e^{\gamma(x)} =e^{\gamma_{0}}$), this yields the coefficient in the
$\Lambda$ term in (\ref{eq:tdeptxxbound}). Therefore, if the profile
curve and its first derivative match their corresponding vacuum
functions at the boundary, $T^{x}_{\;x}$ will vanish there, satisfying
the junction condition (\ref{eq:syngcond2}). At this point, all
junction conditions are met.

\qquad As with the static case, there are many ways to enforce
continuity of $\p$ and $\p_{,x}$ at the boundary. Comparing the matter
profile curve, (\ref{eq:prof}) and (\ref{eq:dprof}), with the vacuum
(\ref{eq:tdepvac}) one may set, for example,
\begin{subequations}
\begin{align}
A^{2}=&\left\{\left[\frac{D-2}{\Lambda}-(x_{b}-x_{0})^{2}\right]^{1/2}
-
P_{0} \right\}\frac{e^{-h^{2}(x_{b})}}{x_{b}^{2n}}, \\
h(x)=&\sqrt{2xF_{0}+ (x-x_{b})^{2p}u(x)},
\end{align}
\end{subequations}
with
\begin{equation}
F(x_{b}):=-\frac{(x_{b}-x_{0})
\left[\frac{D-2}{\Lambda}-(x_{b}-x_{0})^{2}\right]^{-1/2}}{2\left\{
\left[\frac{D-2}{\Lambda}-(x_{b}-x_{0})^{2}\right]^{1/2} -P_{0}
\right\}} -\frac{n}{x_{b}} .
\end{equation}

\qquad Now we wish to investigate the WEC. Consider first the tension
generated condition (\ref{eq:econd2a}) with (\ref{eq:econd1}).
Positivity of $-T^{t}_{\;t}$ has already been established and so we
next concentrate on negativity of $T^{x}_{\;x}$.

\qquad Near the throat we have
\begin{eqnarray}
8\pi T^{x}_{\;x}&\approx& -\frac{1}{P_{0}^{2}} +\Lambda
-\frac{D-4}{4}e^{-\gamma(0)} \left[\frac{1}{2}(D-7)
\left(\frac{\at_{,t}}{\at}\right)^{2} + 2 \frac{\at_{,t,t}}{\at}
\right] -\frac{(D-5)(D-4)}{2\at} \nonumber \\
&&+\frac{D-4}{4}e^{-\gamma(0)}x_{b}^{l}g(0)\left[\frac{1}{2}(D-7)
\left(\frac{\at_{,t}}{\at}\right)^{2} + 2 \frac{\at_{,t,t}}{\at}
\right] x  \nonumber \\
&&+\frac{2n}{P_{0}}x_{b}^{l}g(0)A^{2}e^{h^{2}(0)} x^{2n-1}
+\mathcal{O}(x^{2} \mbox{ or greater}), \label{eq:tdeptxxthrt}
\end{eqnarray}
where we have used (\ref{eq:tdepgx}). Therefore, in the vicinity of
the throat, a negative radial pressure, or tension, may easily be
enforced.

\qquad Exactly at the throat, the first energy condition
(\ref{eq:econd2a}) yields
\begin{equation}
\frac{(D-4)}{2}\left\{ e^{-\gamma(0)} \left[\frac{1}{2}
\left(\frac{\at_{,t}}{\at}\right)^{2} -\frac{\at_{,t,t}}{\at} \right]
-\frac{e^{-\gamma(0)/2}}{2}x_{b}^{l}\,|g(0)
\frac{\at_{,t}}{\at}|\right\} \geq 0. \label{eq:tdepecondthrt}
\end{equation}
Although it seems likely that the above inequality can be satisfied, it
is actually \emph{impossible}. This is because the scale factor is
dictated by the vacuum solution and the term in square brackets is,
from (\ref{eq:aeq}), a \emph{negative constant}. Note that this case
is actually worse than the static case since, for the static case, the
WEC is respected at the throat.

\qquad We next turn our attention to the energy condition
(\ref{eq:econd2b}). Unlike the previous condition, this condition is
generated by positive pressure instead of tension. It is therefore
more realistic from an astrophysical point of view.

\begin{figure}[h!]
\begin{center}
\includegraphics[bb=97 101 426 645, clip, scale=0.68, keepaspectratio=true]{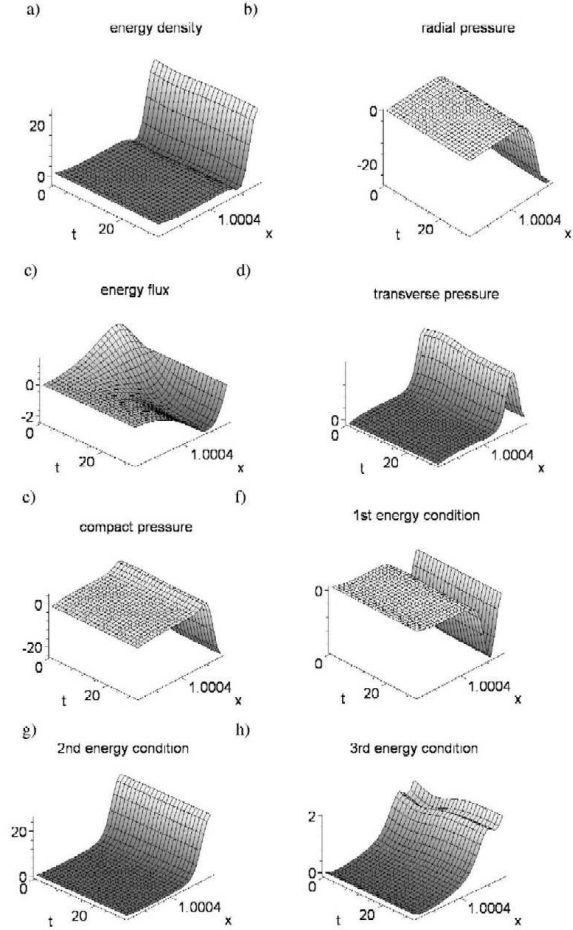}
\caption{{\small An example of a quasi-static wormhole. The graphs
indicate: a) $-T^{t}_{\;t}$, b) $T^{x}_{\;x}$, c) $T^{t}_{\;x}$, d)
$T^{\theta}_{\;\theta}$, e) $T^{A}_{\;A}$, f)
$T^{x}_{\;x}-T^{t}_{\;t}-2e^{\gamma(x)/2}[1+\p_{,x}^{2}]^{-1/2}|T^{t}_{\;x}|$,
g) $T^{\theta}_{\;\theta}-T^{t}_{\;t}$, h)$T^{A}_{\;A}-T^{t}_{\;t}$.
The functions are plotted from the throat to the boundary ($x$-axis)
and over a period of time ($t$-axis). The vertical scales differ for
each plot.}} \label{fig:tdepecnd}
\end{center}
\end{figure}

\qquad It is simple to check that the WEC at the throat \emph{cannot}
be satisfied for either (\ref{eq:econd2b}) or (\ref{eq:econd2c}). To
see this recall that both these conditions require $T^{x}_{\;x}\geq 0$
and $T^{t}_{\;t}\leq 0$. Therefore, the linear combination of $8\pi
\left[T^{x}_{\;x}-T^{t}_{\;t}\right]$ must be non-negative. Forming
this combination using (\ref{eq:tdepTttthroat}) and
(\ref{eq:tdeptxxthrt}) gives:
\begin{equation}
8\pi\left[T^{x}_{\;x}-T^{t}_{\;t}\right]_{|x=0}
=\frac{D-4}{2}e^{-\gamma(0)} \left\{\frac{1}{2}
\left[\frac{\at_{,t}}{\at}\right]^{2}-\frac{\at_{,t,t}}{\at}\right\}.
\end{equation}
As with the previous condition, the quantity in braces is a
negative constant. Again, unlike the four dimensional and static
cases, a WEC respecting throat is \emph{forbidden}. Even at
points away from the throat, the time dependent terms in
$T^{x}_{\;x}-T^{t}_{\;t}$ conspire to yield a negative constant.
Therefore, since all energy conditions imply the positivity of
some expression which possesses the linear combination
$T^{x}_{\;x}-T^{t}_{\;t}$ (a necessary but not sufficient
condition), the time dependence tends to impede rather than
favour the WEC. This is illustrated in the specific example of
figure \ref{fig:tdepecnd}. Figure \ref{fig:tdepecnd}a and b shows
that it is possible to have almost everywhere positive energy
density and negative radial pressure (as required for
(\ref{eq:econd2a})). The graphs in figures \ref{fig:tdepecnd}f-h
are plots of the various energy conditions (see figure caption).
Positive values indicate WEC respecting regions. These figures
show that it is much more difficult than the static case to have
WEC respecting regions in the star.

\qquad Studies have shown that if one introduces time dependence
in the four dimensional case, the WEC may be respected for
arbitrarily long periods of time \cite{ref:kar} - \cite{ref:li}.
This is due to the fact that the time dependence may be chosen in
such a way as to yield the desired result. The above analysis
differs in that the matter field does not possess infinite
spatial extent. That is, the global picture is constructed via
the introduction of the matter-vacuum boundary. When the models
are patched to the vacuum the time dependence of the extra
dimensions is now fixed by the vacuum and cannot be chosen
arbitrarily. It is this restriction which tends the solution
towards WEC violation.

\section{Concluding remarks}
\qquad In this paper we considered static and quasi-static
spherically symmetric wormholes with and arbitrary number of
extra compact dimensions. The matter models were patched to the
vacuum. The vacuum equations were solved and both static and
time-dependent vacua were found. Serious restrictions are placed
on the wormhole due to the presence of the vacuum-matter boundary
and extra dimensions. If one wishes to respect the WEC at the
throat in the static case, the extra dimensions place a
restriction on the radial size of the wormhole throat. Its radius
in the uncompact dimension must be of the order of the size of
the compact dimension. The only exception is for $D=5$. For the
quasi-static case, the WEC \emph{cannot} be satisfied at the
throat even though seemingly arbitrary time-dependent terms are
present in the solution. The WEC \emph{can}, however, be
satisfied throughout most of the stellar bulk in the static case
and in isolated regions of the quasi-static star.

\section*{Acknowledgements}
The authors are grateful to their home institutions for various
support during the production of this work. Also, A. DeB. thanks
the S.F.U. Mathematics department for kind hospitality. We would
also like to thank the anonymous referee for useful comments.

\newpage
\linespread{0.6}
\bibliographystyle{unsrt}

\end{document}